\documentclass[twocolumn,showpacs]{revtex4-1}
\usepackage{amsmath}    % need for subequations
\usepackage{amssymb}
\begin{document}

\title{Comment on the higher derivative Lagrangians in relativistic theory}
\author{Mathieu Beau}
 \affiliation{School of Theoretical Physics, Dublin Institute for Advanced Studies, 10 Burlington Road, Dublin 4}
 \email{mbeau@stp.dias.ie}  

\date{\today}

\begin{abstract}
We discuss the consequences of higher derivative Lagrangians 
of the form $\alpha_1 A_{\mu}(x)\dot{x}^\mu$, $\alpha_2 G_{\mu}(x)\ddot{x}^\mu$, $\alpha_3 B_{\mu}(x)\dddot{x}^\mu$, $\alpha_4 K_{\mu}(x)\ddddot{x}^\mu$, 
$\cdots$, $U_{(n)\mu}(x)x^{(n)\mu}$ in relativistic theory. 
After establishing the equations of the motion of particles in these fields, 
we introduce the concept of the generalized induction principle assuming the coupling between the higher fields $U_{(n),\mu}(x),\ n\geq1$
with the higher currents $j^{(n)\mu}=\rho(x)x^{(n)\mu}$, where $\rho(x)$ is the spatial density of mass or of electric charge.
In addition, we discuss the analogy of the field $G_\mu(x)$ with the gravitational field 
and its inclusion in the general relativity framework in the last section.
This letter is an invitation to reflect on a generalisation of the concept of inertia 
and we also discuss this problem in the last section.
\end{abstract}

\maketitle

\section{Higher derivative Lagrangians and dynamic equations of a particle}

Ostrogradsky introduced the idea of higher derivative Lagrangians in classical mechanics \cite{MecaG0},
and there is a series of articles published about this topic, see \cite{MecaG1},\cite{MecaG2},\cite{MecaG3},\cite{MecaG4},\cite{GeEM}. 
However, to my knowledge, there is no article dealing with relativistic higher derivative Lagrangians of this type:
\begin{multline}\label{L}
\widetilde{L}(\dot{x},\ddot{x},\cdots,x^{(n)})\\
=\alpha_1 A_{\mu}(x)\dot{x}^\mu+\alpha_2 G_{\mu}(x)\ddot{x}^\mu+\cdots+\alpha_n U_{(n)\mu}(x)x^{(n)\mu} 
\end{multline}
where $x^{(n)}(s)\equiv d^n x(s)/ds^n$ are the $n$-derivatives of the position ($ds=cd\tau$, where $\tau$ is the proper time),
and $U_{(n)\mu}(x),\ n=1,2,3,..$ are the generalized vectorial fields coupling linearly with the $n$-derivatives.
Here we denote the field $U_{(1)\mu}(x)=A_\mu(x)$ to refer to the electromagnetic potential.
Also we denote $U_{(2)\mu}(x)=G_\mu(x)$ because of the analogy with the geodesic equations that we will see in the equation (\ref{EqGeodesique}). 
From (\ref{L}), we set the action:
\begin{equation*}
S=\int ds L_0(\dot{x})+\int ds  \widetilde{L}(\dot{x},\ddot{x},\cdots,x^{(n)}) \ ,
\end{equation*}
where $L_0(\dot{x})\equiv\frac{mc^2}{2}\dot{x}_\mu\dot{x}^\mu$.
We will not give an explicit general dynamic theory for a given $n$, we consider only $n=2$ for the moment and we will discuss the general case later. 
An integration by part for $n=2$ gives the equivalent action \cite{MecaG3}:
\begin{equation*}
\widetilde{S}=\alpha_1\int ds A_{\mu}(x)\dot{x}^\mu - \alpha_2 \int ds\ \partial_\nu G_{\mu}\dot{x}^\mu\dot{x}^\nu\ ,
\end{equation*}
and one can see that the first part of the action $\widetilde{S}$ is similar to the electrodynamic action
whereas the second part is similar to the gravitational action.
Indeed, from the generalized Euler-Lagrange equations 
(see \cite{MecaG0},\cite{MecaG1},\cite{MecaG2},\cite{MecaG4}):
\begin{eqnarray}\label{Euler-Lagrange}
\frac{d^2}{ds^2}(\frac{\partial L}{\partial \ddot{x}^\mu})-
\frac{d}{ds}(\frac{\partial L}{\partial \dot{x}^\mu})+\frac{\partial L}{\partial x^\mu}=0\ ,
\end{eqnarray}
with $L=L_0+\widetilde{L}$, we get:
\begin{eqnarray}\label{EqGeodesique}
mc^2\eta_{\mu\nu}\ddot{x}^\nu-\alpha_2\varepsilon_{\mu\nu}\ddot{x}^\nu-\alpha_2\Delta_{\mu\nu\sigma}\dot{x}^\nu\dot{x}^\sigma
=-\alpha_1 F_{\mu\nu}\dot{x}^\nu
\end{eqnarray}
where $\varepsilon_{\mu\nu}$ and $\Delta_{\mu\nu\sigma}$ are:
\begin{eqnarray}\label{DefTenseurs}
\begin{array}{ll}
\varepsilon_{\mu\nu}&\equiv\partial_\mu G_\nu + \partial_\nu G_\mu \label{g}\\
\Delta_{\mu\nu\sigma}&\equiv\partial_\nu\partial_\sigma G_\mu 
=\frac{1}{2}(\partial_\nu \varepsilon_{\mu\sigma}+
\partial_\sigma \varepsilon_{\mu\nu}-\partial_\mu \varepsilon_{\nu\sigma}) \label{gamma1}\end{array}   
\end{eqnarray}
and where $F_{\mu\nu}=\partial_\mu A_\nu - \partial_\nu A_\mu$.
One sees the analogy with the equations of the motion of a charged particle in a gravitational field and in an electromagnetic field.
However, the fixed metric (or \textit{background} metric) is Minkowskian. 
Then, the quadrivector field $G_\mu(x)$ can be seen as a \textit{displacement vector field} 
and $\varepsilon_{\mu\nu}$ can be viewed as an infinitesimal strain tensor
by analogy with the deformation theory of a continuous medium \cite{MMC}.

Now, let us take $n=3$, we denote $U_{(3)\mu}\equiv B_\mu$. From
\begin{eqnarray}\label{Euler-Lagrange3}
-\frac{d^3}{ds^3}(\frac{\partial L}{\partial \dddot{x}^\mu})+\frac{d^2}{ds^2}(\frac{\partial L}{\partial \ddot{x}^\mu})-
\frac{d}{ds}(\frac{\partial L}{\partial \dot{x}^\mu})+\frac{\partial L}{\partial x^\mu}=0
\end{eqnarray}
one has:
\begin{multline}\label{EqGeodesique3}
mc^2\eta_{\mu\nu}\ddot{x}^\nu+\alpha_3 H_{\mu\nu}\dddot{x}^\nu-\alpha_3\Upsilon_{\mu\nu\sigma\rho}\dot{x}^\nu\dot{x}^\sigma\dot{x}^\rho\\
-3\alpha_3\varSigma_{\mu\nu\sigma}\ddot{x}^\nu\dot{x}^\sigma
\alpha_2\varepsilon_{\mu\nu}\ddot{x}^\nu-\alpha_2\Delta_{\mu\nu\sigma}\dot{x}^\nu\dot{x}^\sigma
=-\alpha_1 F_{\mu\nu}\dot{x}^\nu 
\end{multline}
where 
\begin{eqnarray}\label{DefTenseurs3}
&& H_{\mu\nu}\equiv\partial_\mu B_\nu - \partial_\nu B_\mu \label{g}
{}\nonumber\\{}&&\varSigma_{\mu\nu\sigma}\equiv\partial_\nu\partial_\sigma B_\mu
{}\nonumber\\{}&&\Upsilon_{\mu\nu\sigma\rho}\equiv\partial_\nu\partial_\sigma\partial_\rho B_\mu 
\end{eqnarray}
We can see that this field generalizes the idea of the electromagnetic field because of the antisymmetry of $H_{\mu\nu}$.
However, in (\ref{EqGeodesique3}) there are some other fields, similar to $\Delta_{\mu\nu\sigma}$, coupling with the combinations
of the odd derivatives of $x^{\mu}$, i.e. $\ddot{x}^\nu\dot{x}^\sigma$ and $\dot{x}^\nu\dot{x}^\sigma\dot{x}^\rho$.

We can also discuss the higher derivative terms.
For $n=4$, we denote the field $K_\mu(x)\equiv U_{(4)\mu}(x)$. 
The dynamic equations have a similar structure to the one we get for $G_\mu(x)$ (i.e. for $n=2$).
As it has been shown that in the non-relativistic theory \cite{MecaG3}, we notice that the Lagrangian $\alpha_4 x_\mu\ddddot{x}^\mu$ 
is equivalent to this Lagrangian $\alpha_4 \ddot{x}_\mu\ddot{x}^\mu$ 
and the quantity $\alpha_2\dot{x}^2+\alpha_4\ddot{x}_\mu\ddot{x}^\mu-2\alpha_4\dot{x}_\mu\dddot{x}^\mu$ 
could be interpreted as a more \textit{general kinetic energy} \cite{MecaG4}.
Here the problem is similar: the Lagrangian $\alpha_4 K_{\mu}(x)\ddddot{x}^\mu$ 
is equivalent to $\alpha_4\partial_{\mu}K_\nu \ddot{x}^\mu\ddot{x}^\nu
+\alpha_4 \partial_{\sigma}\partial_{\mu}K_\nu \ddot{x}^\mu\dot{x}^\nu\dot{x}^\sigma$ and
so is obviously more complicated than the equivalent Lagrangian that we get for the special case $K_\mu(x)=x_\mu$.

To finish this section, let us now consider the generalized fields $U_{(n)\mu}(x),\ n\geq1$. 
By the generalized Euler-Lagrange equations: 
$$\sum_{k=0}^{n}(-1)^k \frac{d^k}{ds^k}\frac{\partial L}{\partial x^{(k)\mu}}=0$$
we get terms of the form $\partial_{\mu_{1}}\cdots \partial_{\mu_{p}}U_{(n)},\ p=1,\cdots,n$ 
multiplied by the combination of the derivatives $x^{(l_1)\mu_1}x^{(l_2)\mu_2}\cdots x^{(l_p)\mu_p}$, $\sum_{j=1}^{p}l_j=n$.   
One can see that the ``even'' $n$-fields are analogous to the "gravitational field" 
as the ``odd'' $n$-fields are to the "electromagnetic field",
since, in the dynamic equations, the derivatives $x^{(n)\mu}$ are multiplied by the 
symmetric (if $n$ is even) / antisymmetric (if $n$ is odd) part of the 
first derivative of the field:
$$(\partial_{\mu} U_{\nu}+(-1)^n\partial_{\nu}U_{\mu})x^{(n)\nu}\ ,$$
here we denote the $n$-field by $U_\mu$.
We will see the consequences of this remark in the next section.

\section{General fields hypothesis}

The point is to relate the possible existence of these fields with a generalized induction phenomenon.
We will suppose that there exists new physical couplings with \textit{higher currents}, yet unkown, 
and we will generalize the electromagnetic field theory.

\subsection{Construction of the $n=2$-field equations by analogy with the vectorial electromagnetic field}

In the Lagrangian (\ref{L}) for $n=2$ we notice that the field $G_\mu$ is coupled with the acceleration of the particle
as the field $A_{\mu}$ is coupled with the velocity of the particle.
By analogy with the construction of the electromagnetic field theory, we suggest the following field equations:  
\begin{equation}\label{gpe2}
\partial_\mu \varepsilon^{\mu\nu}(x) = -\kappa j^{(2)\nu}(x)\ ,
\end{equation}
where the acceleration current density $j^{(2)\nu}$ (generally non-conserved) is: 
\begin{equation}\label{QuadridensitAccRG}
j^{(2)\nu}(x)\equiv \rho_m(x)c^2\frac{du^\nu}{ds}\ ,
\end{equation}
where $\rho_m(x)$ is the density of particles and $\frac{du^\nu}{ds}$ is the 4-acceleration. 
Let us rewrite the coupling constant $\kappa$ as follows 
$$\kappa=\frac{8\pi G \lambda^2}{c^4}$$ 
where $\lambda$ has the dimension of a length.

To complete the system of field equations, we need ten equations: 
\begin{equation}\label{gpe1}
\partial_\sigma \partial^\sigma \varepsilon_{\mu\nu}+\partial_\mu \partial_\nu \varepsilon_{\sigma}^{\sigma}
=\partial_\mu \partial^\sigma \varepsilon_{\sigma\nu}+\partial_\nu \partial^\sigma \varepsilon_{\sigma\nu}\ ,
\end{equation}
The equations (\ref{gpe1}) are analogous to the compatibility equations 
for the strain tensor in the three-dimensional non-relativistic theory of deformation of continuous media \cite{MMC}.

Hence we get the following wave equations:
\begin{eqnarray}\label{PropaChampsPhys0}
\Box \varepsilon_{\nu\sigma}(x)+\partial_{\nu}\partial_\sigma \varepsilon_\mu^\mu(x)=-\kappa \xi_{\nu\sigma}^{(2)}(x) 
\end{eqnarray}
with $\xi^{(2)}_{\nu\sigma}(x)\equiv\partial_\sigma j^{(2)}_\nu(x)+\partial_\nu j^{(2)}_\sigma(x)$ .
Also, the trace of $\varepsilon_{\mu\nu}$ follows this equation 
\begin{eqnarray}\label{Tracen2}
\Box \varepsilon_{\mu}^\mu(x)=-\kappa\partial_\mu j^{(2)\mu}(x)\ , 
\end{eqnarray}
this means that $\varepsilon_\mu^\mu$ is a non-massive scalar field.
In the linear theory of the deformation of a continuous medium, 
the trace of the strain tensor is interpreted as the contraction/dilation of the volume \cite{MMC}.
Here we observe a similar property and by the equation (\ref{Tracen2})
we conclude that the relativistic deformation of the volume of the \textit{four dimensional continuous medium} 
is related to the non-conservation of the current $j^{(2)}$.

\subsection{Generalisation to the $n$-field equations}

With respect to the analogy that we have already discussed above between the even fields and the gravitational field 
and between the odd fields with the electromagnetic field and using the field theory developed for the case $n=2$,
we can construct the general field theory via a more general induction principle.
Following this rule, we rewrite the constants in (\ref{L}) as $\alpha_{2n}=mc^2(\lambda_{n})^{2n-2}$ 
and $\alpha_{2n-1}=\frac{(\xi_{n})^{2n-2}}{c^{2n-2}},\ n\geq1$,
where $\lambda_n$ and $\xi_n$ are fundamental 'length' constants and $G$ is the gravitational constant. 

It comes naturally that for so-called \textit{gravitational type fields} $U_{(2n)}\equiv G_{(n)},\ n\geq1$,
the coupling has the form:
\begin{equation}
-\frac{8\pi G}{c^2}\frac{(\lambda_{n})^{2n}}{c^{2n}}G_{(n)\mu}(x)j^{(2n)\mu}(x)\ ,
\end{equation}
whereas for the so-called \textit{electromagnetic type fields} $U_{(2n-1)}\equiv A_{(n)},\ n\geq1$, the coupling has the form:
\begin{equation}
\mu_0\frac{(\xi_{n})^{2n-2}}{c^{2n-2}}A_{(n)\mu}(x)j^{(2n-1)\mu}(x)\ ,
\end{equation}
where $A_{(n)\mu}$ has the dimension of $\mathrm{N.A^{-1}}$ ($\mathrm{N}$ is the Newton and $\mathrm{A}$ the Amp\`ere),
$\mu_0$ is the vacuum permeability ($\mu_0=4\pi\times 10^{-7}\ \mathrm{N.A^{-2}}$),
and where $G_{(n)\mu}$ has the dimension of a length.
The generalized currents for $n=1,2,3,..$ are constructed as follows:
\begin{equation}
j^{(n)\nu}\equiv 
\left\{\begin{array}{ll}
\rho_m(x)\frac{d^{n}x^\nu}{d\tau^{n}},\ \mathrm{if\ n\ is\ even} \\
\rho_e(x)\frac{d^{n}x^\nu}{d\tau^{n}},\ \mathrm{if\ n\ is\ odd}
\end{array}\right.
\end{equation}
where $\rho_m(x)$ is the mass density and $\rho_e(x)$ the electric charge density.
Similar to (\ref{gpe2}), we construct an $(2n-1)$-order linear differential theory to relate the sources and the fields:
\begin{equation}\label{gpe2n}
O^{(n)}_\mu(\lambda_n) \epsilon_{(n)}^{\mu\nu}(x) = - \frac{8\pi G}{c^2}\frac{\lambda_n^{2n}}{c^{2n}} j^{(2n)\nu}(x),\ n\geq1\ ,
\end{equation}
and
\begin{equation}\label{gpe2n}
Q^{(n)}_\mu(\xi_n) f_{(n)}^{\mu\nu}(x) = - \mu_0\frac{\xi_n^{2n-2}}{c^{2n-2}} j^{(2n-1)\nu}(x),\ n\geq1\ ,
\end{equation}
where $O^{(n)}_\mu(\lambda_n)$ and $Q^{(n)}_\mu(\xi_n)$ are two $(2n-1)$-order differential operators 
and where $\varepsilon_{(n)}^{\mu\nu}(x)\equiv \partial^\mu G_{(n)}^\nu+\partial^\nu G_{(n)}^\mu$
and $f_{(n)}^{\mu\nu}(x)\equiv \partial^\mu A_{(n)}^\nu-\partial^\nu A_{(n)}^\mu$. 

From those rules we could obtain similar wave equations to (\ref{PropaChampsPhys0}) and (\ref{Tracen2})
but with a higher order differential operator $(\lambda_n)^{2k}\underbrace{\Box\Box\cdots\Box}_{k\ \mathrm{times}},\ k=1,\cdots,n$. 
For example, for the $4$-field we can take $O^{(4)}_\mu(\lambda)=(\lambda^2\Box+1)\partial_\mu$ and
then we get the wave equation for the trace of the tensor $\zeta_{\mu\nu}\equiv \partial_\mu K_\nu+\partial_\nu K_\mu$: 
\begin{eqnarray}\label{Tracen4}
\left(\lambda^2\Box+1\right)\Box \zeta_{\mu}^\mu(x)=- \frac{8\pi G\lambda^4}{c^6} \partial_\mu j^{(4)\mu}(x)\ ,
\end{eqnarray} 
and then $\zeta_\mu^\mu$ is a massive scalar field. 

Notice that for the electromagnetic type fields, the choice of the fields $A_{(n)\mu}$ 
(due to the antisymmetry of the fields $f_{(n)}^{\mu\nu}$) is not unique
whereas for the gravitational type fields all of the components of the fields $G_{(n)\mu}$ are physical.

This project is at an early stage and the construction of these operators $O^{(n)}_\mu$ and $Q^{(n)}_\mu$ 
has to be understood, even for $n=2$.

\subsection{Unitary fields}

Physically, we can understand the \textit{generalized vectorial fields theory} 
as perturbative corrections of the first order theory 
(i.e. $A_{(2)\mu}\equiv B_\mu(x)$ is a correction of the Minkowskian theory of Electromagnetism field $A_{(1)\mu}\equiv A_\mu$).
Therefore, it is natural to unify the gravity type fields as well as the electromagnetic type fields.
Then, we construct the dimensionless unification constants: 
$$\gamma_{jl}=\frac{\lambda_j}{\lambda_l},\ \theta_{jl}=\frac{\xi_j}{\xi_l},\ j,l=1,2,3,\cdots$$
where the constants $\lambda_n,\ n\geq1$ and $\xi_{n},\ n\geq1$,
were introduced in section II.B.

For example, if we suppose that $A_\mu(x)=B_\mu(x)$, we get the coupling:
$$\mu_0 A_{\mu}(x)\left( j^{(1)\mu}(x)+\frac{\xi^2}{c^2}j^{(3)\mu}(x)\right)$$
where we put $\xi_2=\xi$ (remind that $\alpha_1=1$ and $\alpha_3=\xi_2^2/c^2$).
Phenomenologically, this means that for an electric circuit with an intensity of this type $I(t)=I_0 t^2/\tau^2$,
where $\tau$ is a time constant, the third derivative of the electrons in the current is non-zero 
(this kinematic quantity is called the \textit{Jerk}, see \cite{Jerk})
and so that the electromagnetic field would be modified by the \textit{jerk current} $j^{(3)}$. 
We mention that in the generalized theory of Electrodynamics \cite{GeEM}
the relation with the higher derivatives currents has not been suggested.

Similarly, we can construct the unified coupling for the even fields.
for example, if we suppose that $G_{\mu}(x)=K_\mu(x)\equiv G_{(2)\mu}(x)$, 
we can reconstruct the unified coupling for $n=2,4$ in the following way:
$$-\frac{8\pi G}{c^2}\frac{\lambda^{2}}{c^{2}}G_{\mu}(x)\left( j^{(2)\mu}(x)+\gamma^4\frac{\lambda^2}{c^2}j^{(4)\mu}(x)\right)$$
where we put $\lambda_1=\lambda,\ \gamma=\gamma_{21}$ and where we introduce the effective current
$\widetilde{j}^{(2)\mu}=j^{(2)\mu}+\gamma^4\frac{\lambda^2}{c^2}j^{(4)\mu}$.
So the effective deformation $\epsilon_{\mu\nu}$ to the Minkowski metric
is also induced by the second derivative of the acceleration of the particles moving in the space.

\section{Comments}

\begin{itemize}
 \item Microscopic Physics and generalized currents
\end{itemize}

The effect of gravitation at the microscopic scale is not yet well known. 
We wonder if the current $j^{(2)}$ of the acceleration of masses will be significant.
There is no current proof that gravitation can be viewed as a metric field at this scale.
It might also be a challenge to see whether the higher derivative fields play a role in particle physics.
We gave the interpretation of the $2n$-fields (ex $G_\mu$ and $K_\mu$) and between the $(2n-1)$-fields (ex $A_\mu$ and $B_\mu$) 
as the generalisation of the gravitational and electromagnetic field for higher currents.
Hence, formulating a perturbative quantum field theory including these fields is an open question.
$\ $\\

\begin{itemize}
 \item Generalized kinematic model and special relativity theory
\end{itemize}

After introducing the higher derivative fields and the higher currents, 
we naturally wonder if it is possible to extend the Lorentzian kinematic theory
to a more general inertial concept where the higher derivatives of the quadri vector position
appear in the free Lagrangian. 
For example, following the discussion in section I, we propose the following Lagrangian:
\begin{equation}\label{Lolambda}
L_{\lambda;\alpha}(\dot{x},\ddot{x})=\frac{mc^2}{2}\left(\dot{x}_\mu\dot{x}^\mu + \alpha\lambda^2\ddot{x}_\mu\ddot{x}^\mu\right)\ , 
\end{equation}
where $m$ is the mass of the particle, $\lambda$ is a universal constant and $\alpha=\pm1$.
For this model the free motion is not determined by $\ddot{x}^\mu=0$
but by the equations: 
\begin{equation}\label{EqLolambda}
\ddddot{x}^\mu=-\frac{\alpha}{\lambda^2}\ddot{x}^\mu
\end{equation}
independently to the mass of the particle. 
We question whether the model (\ref{Lolambda}) is consistent with the special theory of relativity. 
One problem concerns the simultaneity. 
Consider an observer in an inertial reference frame, i.e. $\ddot{x}_\mu(s)=0$. 
If we imagine that a photon is emitted from an accelerated massive particle 
(the equation of motion are given by (\ref{Lolambda}) with the initial conditions $\dddot{x}_\mu(0)=0$ and $\ddot{x}_\mu(0)=a_\mu$),
is the photon accelerated with respect to the inertial reference frame observing the motion of the particle 
or does its velocity remain constant and equal to $c$ ? 
Behind this question is a more fundamental problem of the equivalence between the accelerated reference frames. 
This requires further study.  
$\ $\\

\begin{itemize}
 \item Strain and stress tensor in General Relativity
\end{itemize}

We will discuss only the $n=2$-field model of the section II.A and we consider $\alpha_{n\geq3}=0$.
In a future project, it will be interesting, as a first step, to formulate the $G_\mu$-field theory 
as a perturbative gravitational field arising from the metric field
and to then to look at the possible effects for particle physics and/or cosmology.

We will give here the idea to construct the covariant strain/stress field theory.
Let us consider the covariant derivative for a Riemannian metric space $\varepsilon_{\mu\nu}=D_\mu G_\nu+D_\nu G_\mu$. 
Now, we construct a stress tensor
\begin{equation}\label{Sigma}
\sigma_{\mu\nu}(x)=\rho_G c^2\varepsilon_{\mu\nu}(x)
\end{equation}
where $\rho_G c^2=\frac{c^4}{8\pi G \lambda^2}$ is the density of energy constant,
and analogous to the Young modulus for an isotropic medium \cite{MMC}.
The stress tensor could be added to the Einstein field equations of gravity and then we get the relation:
\begin{equation}\label{Conserv}
D_{\mu}\sigma^{\mu\nu}(x)+D_\mu T^{\mu\nu}(x)=0 
\end{equation}
where $T_{\mu\nu}$ is the energy-momentum tensor in the Einstein field equation.
This equation means that the total energy in the universe is conserved but that the ``visible'' energy can be accelerated. 
This variation of inertia is compensated by the divergence of the stress energy of the continuous medium. 
There is no contradiction with Einstein theory of gravitational fields
and this gives a new perspective on the Mach principle revisiting the ``absolute'' acceleration concept 
as a natural motion in space-time deformed by the matter-energy contained therein. 
We refer the reader to the paper of Einstein on a related topic \cite{Einstein}. 
The relativistic theory of an Aether was discussed several times, see for e.g. \cite{Aether1}, \cite{Aether2}.
In this paper, our hypothesis is different and gives a relativistic theory of the deformation of continuous media
(for which the geometry is still described by the metric field whereas the strain tensor is an additional field).
Then, we could construct a more general stress tensor: 
\begin{equation}
\sigma_{\mu\nu}(x)=C_{\mu\nu\alpha\beta}(x)\varepsilon^{\alpha\beta}(x)
\end{equation}
with the \textit{elasticity tensor} $C_{\mu\nu\alpha\beta}(x)$. 
In fact, a consistent theory should give the equations for the elasticity tensor 
(which gives the mechanical property of the four dimensional continuous medium)
and we think it might be related to the metric tensor as well as its derivatives (i.e. to the Ricci tensor).

\appendix


\begin{thebibliography}{0}%
\makeatletter
\providecommand \@ifxundefined [1]{%
 \@ifx{#1\undefined}
}%
\providecommand \@ifnum [1]{%
 \ifnum #1\expandafter \@firstoftwo
 \else \expandafter \@secondoftwo
 \fi
}%
\providecommand \@ifx [1]{%
 \ifx #1\expandafter \@firstoftwo
 \else \expandafter \@secondoftwo
 \fi
}%
\providecommand \natexlab [1]{#1}%
\providecommand \enquote  [1]{``#1''}%
\providecommand \bibnamefont  [1]{#1}%
\providecommand \bibfnamefont [1]{#1}%
\providecommand \citenamefont [1]{#1}%
\providecommand \href@noop [0]{\@secondoftwo}%
\providecommand \href [0]{\begingroup \@sanitize@url \@href}%
\providecommand \@href[1]{\@@startlink{#1}\@@href}%
\providecommand \@@href[1]{\endgroup#1\@@endlink}%
\providecommand \@sanitize@url [0]{\catcode `\\12\catcode `\$12\catcode
  `\&12\catcode `\#12\catcode `\^12\catcode `\_12\catcode `\%12\relax}%
\providecommand \@@startlink[1]{}%
\providecommand \@@endlink[0]{}%
\providecommand \url  [0]{\begingroup\@sanitize@url \@url }%
\providecommand \@url [1]{\endgroup\@href {#1}{\urlprefix }}%
\providecommand \urlprefix  [0]{URL }%
\providecommand \Eprint [0]{\href }%
\providecommand \doibase [0]{http://dx.doi.org/}%
\providecommand \selectlanguage [0]{\@gobble}%
\providecommand \bibinfo  [0]{\@secondoftwo}%
\providecommand \bibfield  [0]{\@secondoftwo}%
\providecommand \translation [1]{[#1]}%
\providecommand \BibitemOpen [0]{}%
\providecommand \bibitemStop [0]{}%
\providecommand \bibitemNoStop [0]{.\EOS\space}%
\providecommand \EOS [0]{\spacefactor3000\relax}%
\providecommand \BibitemShut  [1]{\csname bibitem#1\endcsname}%
\let\auto@bib@innerbib\@empty
%</preamble>
\end{thebibliography}%


\begin{thebibliography}{5}


\bibitem{MecaG0} M. Ostrogradsky, Mem. Acad. St. Petersburg \textbf{6} (4),385 (1850).

\bibitem{MecaG1} M. Borneas, On a generalization of the Lagrange Function, American Journal of Physics, 
\textbf{27} (4), pp. 265-267 (1959).

\bibitem{MecaG2} M. Borneas, Principle of the Action with Higher Derivatives, Phys.Rev \textbf{186} (5), 1299, (1969).

\bibitem{MecaG3} D. Anderson, Equivalent Lagrangians in generalized mechanics, J. Math. Phys. \textbf{14}, 934 (1973).

\bibitem{MecaG4} C. G. Adler, Why is mechanics based on acceleration?, Philosophy of Science, \textbf{47}, 146-152 (1980).

\bibitem{GeEM} B. Podolsky, A Generalized Electrodynamics: Part I - Non-Quantum, Phys. Rev. \textbf{62}, 68-71 (1942).

\bibitem{MMC} W. M. Lai, D. Rubin, E. Krempl, ``Introduction to Continuum Mechanics'', (Butterworth-Heinemann, Fourth edition, 2010).

\bibitem{Jerk} S. H. Schot, ``Jerk: The time rate of change of acceleration'', Am. J .Phys. \textbf{46}, 1090 (1978).

\bibitem{Einstein} A. Einstein, "Ether and the Theory of Relativity" (1920), in Sidelights on Relativity (Methuen, London, 1922)

\bibitem{Aether1} C. Eling, T. Jacobson, D. Mattingly, ``Einstein Aether Theory'',
in Deserfest, eds. J. Liu, K. Stelle, R. P. Woodard (World Scientific, 2006).

\bibitem{Aether2} T. G. Zlosnik, P. G. Ferreira, G. D. Starkmann, 
Modifying gravity with the Aether: an alternative to Dark Matter, Phys. Rev. D \textbf{75}, 044017 (2007).



\end{thebibliography}
\end{document}